\documentclass[11pt, a4paper]{article}
\pdfoutput=1
\usepackage{jheppub}
\usepackage{amsmath}
\usepackage{graphicx}
\usepackage[config=altsf]{subfig}
\usepackage{feynmf}
\usepackage{dsfont}
\usepackage{array}
\usepackage{slashed}
\usepackage{slashed}
\usepackage{afterpage}
\usepackage{comment}
\usepackage{appendix}
\usepackage{multirow}
\usepackage{dcolumn}
\usepackage{slashed}
\usepackage{soul}
\usepackage{cleveref}
\crefname{figure}{Fig.}{Figs.}

\usepackage{booktabs}    
\usepackage{placeins}
\usepackage{bbm}
\usepackage{amsfonts}
\usepackage{amsmath,amssymb}
\usepackage{mathrsfs}
\usepackage{epsfig}
\usepackage{graphicx}
\usepackage{wrapfig}   

\usepackage{url}
\usepackage{array}
\usepackage{hyperref}
\usepackage{float}
\usepackage{pstricks}
\usepackage{color}
\usepackage{multirow}
\usepackage{multicol}
\usepackage{lipsum}
\usepackage{enumitem}
\usepackage{ctable}

\usepackage{pgfkeys, pgfsys, pgfcalendar}

\vspace*{3cm}

\title{\bf Effect of anomalous $HHH$ coupling on 
the decay $H\rightarrow Z\,Z^*\rightarrow$ 4 charged leptons}

\author[a]{Pankaj Agrawal,} 
\author[b]{Biswajit Das}

\affiliation[a]{Center for Quantum Engineering, Research and Education (CQuERE), TCG CREST, Salt Lake Sector 5, Kolkata 700091, India}
\affiliation[b]{The Institute of Mathematical Sciences,
IV Cross Road, Taramani, Chennai 600113, India}
\emailAdd{pankaj.agrawal@tcgcrest.org}
\emailAdd{biswajitd@imsc.res.in}

\abstract{ We have computed the electroweak corrections to
$H\rightarrow Z\,Z^*\rightarrow$ 4 charged leptons, including the effect of anomalous $HHH$ coupling in the $\kappa$-framework. The results of this scaling are gauge invariant. We have computed the results for $ H \to e^+ e^- \mu^+ \mu^-$ and  $ H \to e^+ e^- e^+ e^-$ processes. The corrections for the both processes depend on the input parameter scheme. In the $G_F$ scheme, the electroweak
corrections are about $1.26\%$ for the  $ H \to e^+ e^- \mu^+ \mu^-$ and about $0.25\%$ for the  $ H \to e^+ e^- e^+ e^-$ process. However changing the $\kappa$ from
$4$ to $-4$, the corrections vary from less than $1\%$
to about $-6\%$. We have plotted a number of kinematic
distributions. The corrections over most of the phase
space regions are similar. These large corrections can
be used to put a bound on the $HHH$ coupling. This can
help in determining the structure of the Higgs potential.
}

\begin{document}
\maketitle

\flushbottom

\newpage

\section{Introduction}
\label{sec:intro}

The standard model has been enormously successful \cite{conference1,conference2}. A 
long-standing muon $g-2$ anomaly has also been resolved recently \cite{Muong-2:2025xyk}.
However, the Higgs sector of the standard model is still
to be fully established. The self-couplings of the Higgs boson are still to be determined. There are many possibilities for the Higgs sector that are still open \cite{Agrawal:2019bpm}.
In addition, in the absence of any signal
beyond-the-standard-model, it is desirable to compute the radiative corrections to the standard model processes to find any possible discrepancy.
    We computed one-loop electroweak (EW) correction to the $H\rightarrow \nu_e\bar{\nu}_e\nu_\mu\bar{\nu}_\mu$ process in our previous work \cite{Agrawal:2024dbt}, where we studied the effects of anomalous $HHH$ and $ZZWW$ couplings. It is difficult to probe this process at colliders as the final state particles are neutrinos. In this letter, we consider
  the $H\rightarrow e^+e^-\mu^+\mu^-$ and  $ H \to e^+ e^- e^+ e^-$ processes to probe the Higgs couplings at colliders. These decay processes have been explored in \cite{PhysRevD.74.013004,Boselli:2015aha}. There is a monte-carlo package {\tt Prophecy4f} based on the papers \cite{PhysRevD.74.013004,Bredenstein:2006nk} . We have compared our standard model results with this code. There is good agreement. 
  
  In this letter, we have studied the effect of anomalous $HHH$ coupling on the electroweak corrections
  to the $H\rightarrow Z\,Z^*\rightarrow$ 4 charged leptons process. In this process, one can probe $HHH$ and $ZZHH$ couplings as they appear in the one-loop EW virtual diagrams. We use $\kappa$-framework to explore the
  $HHH$ coupling. The scaling of the $HHH$ coupling gives
  gauge invariant results, so we can use $\kappa$-framework.
  The expected bound from ALTAS collaboration search for $b\bar{b}b\bar{b}$ in the final state is $-5.4< \kappa_{HHH}< 11.4$~\cite{PhysRevD.108.052003}. The bound is given at $95\%$($2\sigma$) confidence level. The bound will be more loose at $3\sigma$ confidence level. 
  However, gauge symmetries of the model determines the $ZZHH$ coupling. So we cannot scale this coupling, as it will break the gauge symmetry.

  The $H\rightarrow Z\,Z^*\rightarrow$ 4 charged leptons
  processes get contribution from vertex diagrams that
  involve $HHH$ coupling. In addition, there is a Higgs boson self-energy diagram with $HHH$ coupling. This diagram contributes to the
  wave function renormalization constant of the Higgs boson. We have consistently computed
all the diagrams with $HHH$ coupling to study the impact of this
anomalous coupling.
  
   In the next sections, we discuss the process in detail. In Sec.~\ref{sec:prcs}, we discuss the process and the diagrams
   that we need to compute. In Sec.~\ref{sec:calc_check}, we discuss the  computation techniques, dipole subtraction, renormalization and the anomalous coupling.
   In Sec.~\ref{sec:numr_res}, we have shown the numerical results for the SM prediction and the results with the anomalous coupling effects. In the last section, we present some conclusions.

\section{The Process}
\label{sec:prcs}
We are interested in studying the width of the decay channel $H\rightarrow e^+e^-\mu^+\mu^-$, $H\rightarrow 2e^+2e^-(2\mu^+2\mu^-)$. We have taken the charged leptons as massless particles, so the four electron-type and four muon-type charged lepton final state are the same for this study. From this onwards, we will drop our discussion on $H\rightarrow 2\mu^+2\mu-$ as this process is exactly the same as $H\rightarrow 2e^+2e^-$. 
We calculate the next-to-leading order (NLO) electroweak (EW) corrections to these processes. Higgs trilinear coupling ($HHH$) appears in the one-loop EW virtual diagrams of these processes. This coupling can also appear in counterterm (CT) diagrams.
This coupling can be varied in virtual and CT diagrams appropriately to study the effects on the partial decay width of the Higgs boson.

\begin{figure}[!h]
  \begin{center}
\includegraphics [angle=0,width=0.5\linewidth]{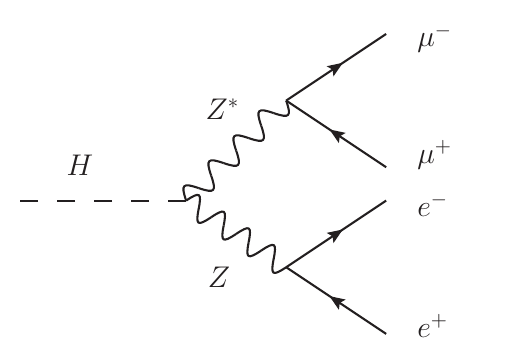}\\
	\caption{The LO Feynman diagram for the decay channel $H \rightarrow e^+e^-\mu^+\mu^-$.  }
	\label{fig:tree_dia}
	\end{center}
\end{figure}

At the leading order, there is only one tree-level diagram for the process $H\rightarrow e^+e^-\mu^+\mu^-$ and two tree-level diagrams for the process $H\rightarrow 2e^+2e^-$. The tree-level diagram for the process $H\rightarrow 2e^+2e^-$ is shown in Fig.~\ref{fig:tree_dia}. There is one extra cross-diagram between electron and muon for the process $H\rightarrow 2e^+2e^-$.  These processes only have intermediate $Z$ bosons that decay to four charged leptons. We allow off-shell intermediate $Z$ bosons in these processes as shown in Fig.~\ref{fig:tree_dia}.

\begin{figure}[!h]
  \begin{center}
\includegraphics [angle=0,width=1\linewidth]{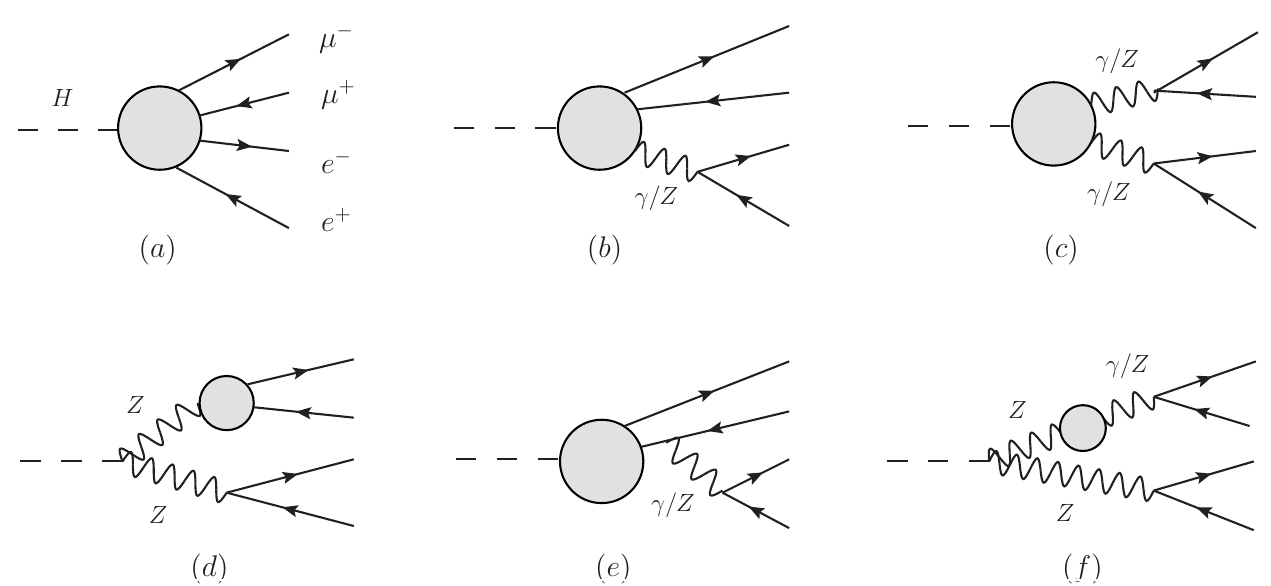}\\
	\caption{Generic class of NLO EW virtual Feynman diagrams for $H\rightarrow e^+e^-\mu^+\mu^-$.  }
	\label{fig:loop_dia}
	\end{center}
\end{figure}

As these process are $1\rightarrow 4$, there are pentagon, box, triangle and bubble-type of virtual diagrams at one loop. There are a total $256$ and $512$ virtual Feynman diagrams at one loop level for the process $H\rightarrow e^+e^-\mu^+\mu^-$ and $H\rightarrow 2e^+2e^-$, respectively.
 The generic classes of one-loop virtual diagrams for the process $H\rightarrow e^+e^-\mu^+\mu^-$ are shown in Fig.~\ref{fig:loop_dia}.
 We will discuss the diagrams for the process for the $H\rightarrow e^+e^-\mu^+\mu^-$ process.
 The diagrams for $H\rightarrow 2e^+2e^-$ will have extra same set of diagrams with crossing between the muon and the electron.
  The generic class of diagrams (a) shown in Fig.~\ref{fig:loop_dia} is the pentagon-type. There are a total of $10$ pentagon-type virtual diagrams in the $H\rightarrow e^+e^-\mu^+\mu^-$ process.  
 The generic class of diagrams (b) displayed in Fig.~\ref{fig:loop_dia} is the box-type diagram. In this generic diagram, the gauge bosons attached to $e^+$ and $e^-$ can be $\gamma$ or $Z$ bosons as depicted in Fig.~\ref{fig:loop_dia}.
  The box loop associated with this generic diagram is attached to Higgs, $\gamma/Z$ bosons, and muons. There is also another similar type of generic diagram (which is not displayed in Fig.~\ref{fig:loop_dia}) where the box loop is attached with Higgs, $\gamma/Z$ bosons, and electrons.
 There are a total of $22$ box-type diagrams in the process $H\rightarrow e^+e^-\mu^+\mu^-$. The generic class of diagrams (c) shown in Fig.~\ref{fig:loop_dia} represents both triangle and bubble-type diagrams. This generic class of diagrams includes the correction to $HZZ$ vertex. It also includes triangle diagrams with loops involving Higgs, $\gamma$, $Z$-boson, and Higgs, $\gamma$, $\gamma$. 
 There are a total of $122$ virtual diagrams that can be represented by the generic diagrams (c).
  The generic class of diagrams (d) shown in Fig.~\ref{fig:loop_dia} is the correction to $Z\mu^+\mu^-$ vertex, and it is a triangle-type diagram. There is also another similar type of diagram for $Ze^+e^-$ vertex, which is not shown in the Fig.~\ref{fig:loop_dia}.
   There are a total of $8$ virtual diagrams that are included in this generic diagram.
  The generic class of diagrams (e) in Fig.~\ref{fig:loop_dia} represents a set of triangle-type diagrams associated with Higgs boson and muons. There is another similar type of generic triangle diagram associated with Higgs boson and electrons, which is not shown in Fig.~\ref{fig:loop_dia}. There are a total of $16$ diagrams that can be represented by this generic diagram.
  The generic class of diagrams (f) in Fig.~\ref{fig:loop_dia} represents the self-energy diagrams. There are $Z\text{-} Z$ and $Z\text{-}\gamma$ type self-energy diagrams as shown in Fig.~\ref{fig:loop_dia}.
   These are the bubble and tadpole-type diagrams. There is also same set of diagrams with the other $Z$ boson propagator, which are not shown in Fig.~\ref{fig:loop_dia}.
    There are a total of $78$ (bubble and tadpole-type) diagrams in this generic class.  
   The generic class of diagrams (c) is very important in this study. It includes a few diagrams as shown in Fig.~\ref{fig:hhh_dia}, where we can study the effect of anomalous $HHH$ coupling.
 We encounter top-quark loop diagrams in the generic class of diagrams (c).

\begin{figure}[!h]
  \begin{center}
\includegraphics [angle=0,width=0.8\linewidth]{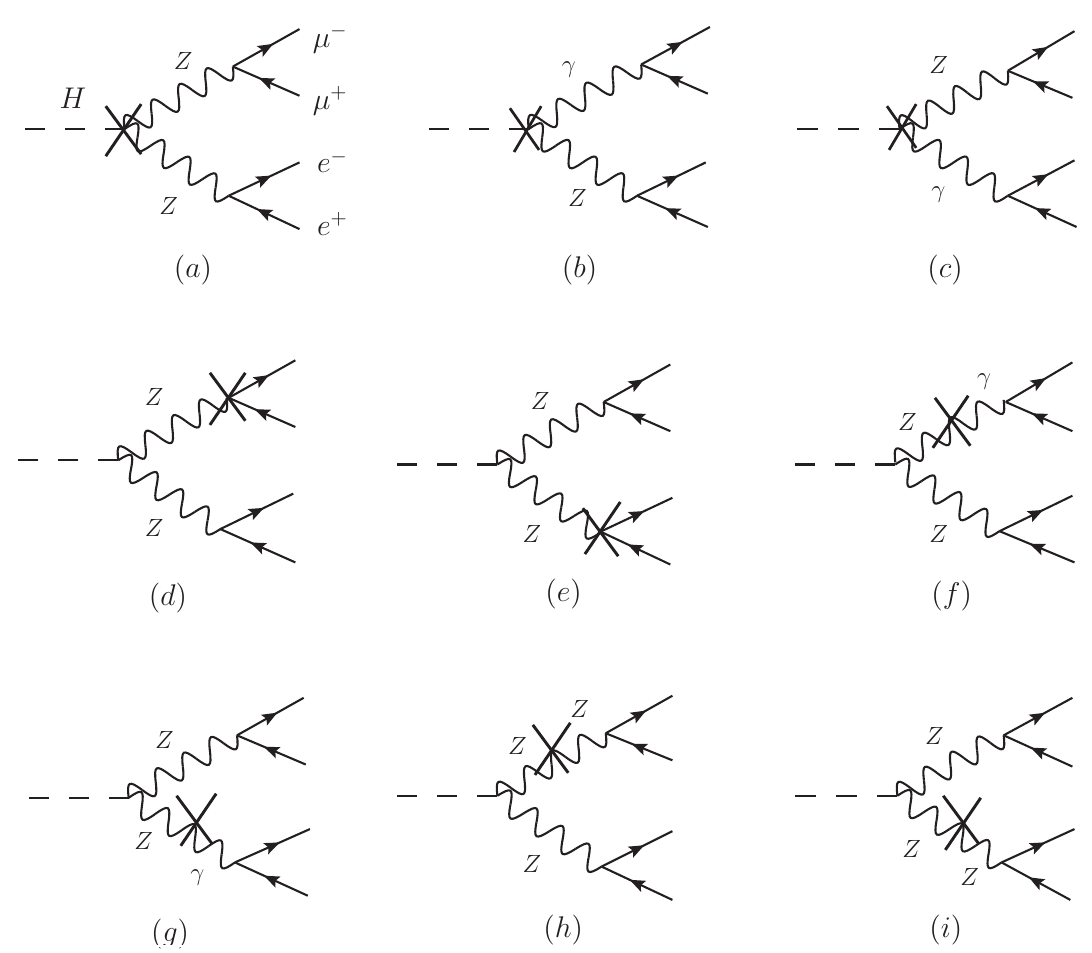}\\
	\caption{Counterterm diagrams for NLO EW correction to $H\rightarrow e^+e^-\mu^+\mu^-$.  }
	\label{fig:ct_dia}
	\end{center}
\end{figure}

We have enlisted the counterterm (CT) diagrams in Fig.~\ref{fig:ct_dia}. As displayed in Fig.~\ref{fig:ct_dia}, there are a total of $9$ CT diagrams in the process $H\rightarrow e^+e^-\mu^+\mu^-$. The diagram (a) is the CT diagram for $HZZ$ vertex correction.
  The diagrams (b) and (c) in Fig.~\ref{fig:ct_dia} are the CT diagrams for the generic loop diagram (c) (in Fig.~\ref{fig:loop_dia}) with the loops associated with Higgs, $\gamma$ and $Z$ boson.
  The diagram (d) in Fig.~\ref{fig:ct_dia} is the CT diagram for $Z\mu^+\mu^-$ vertex correction. Similarly, the diagram (e) in Fig.~\ref{fig:ct_dia} is the CT diagram for $Ze^+e^-$ vertex correction.
  The diagrams (f) and (g) in Fig.~\ref{fig:ct_dia} are the CT diagrams for $Z\text{-}\gamma$ type self-energy diagrams. The diagrams (h) and (i) in Fig.~\ref{fig:ct_dia} are the CT diagrams for $Z\text{-}Z$ type self-energy diagrams.
  CT diagrams are also sensitive to the $HHH$ coupling. As counterterms depend on Higgs, $Z$, and $W$ boson self-energies, these couplings appear in the CT diagrams for these processes.
  
\begin{figure}[!h]
  \begin{center}
\includegraphics [angle=0,width=0.8\linewidth]{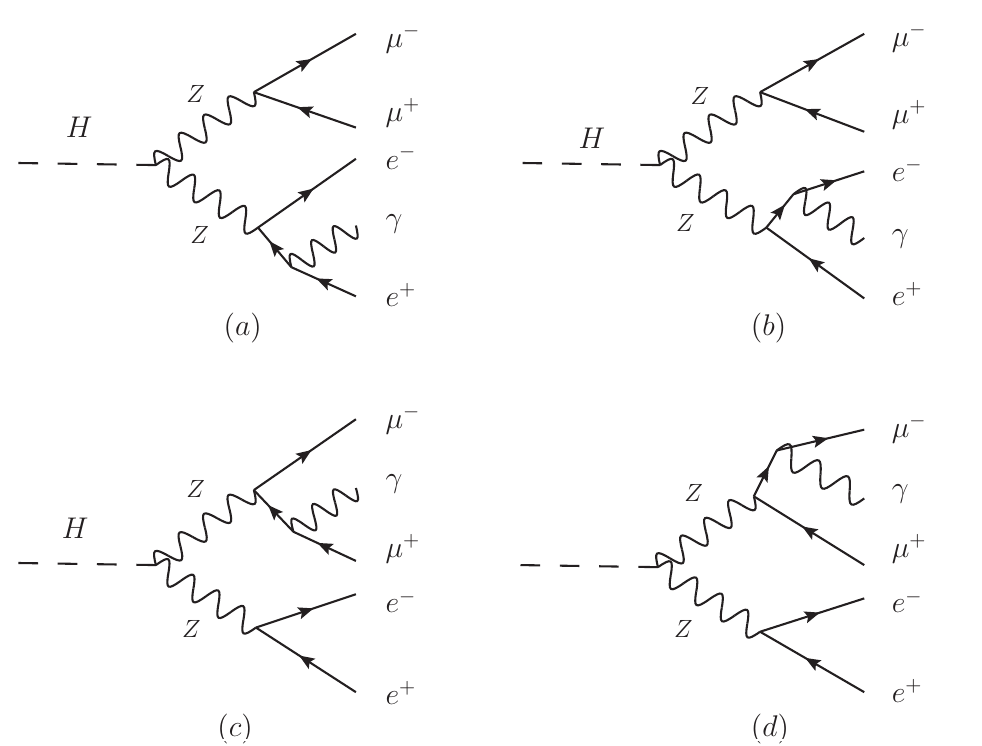}\\
	\caption{Real emission diagrams for NLO EW correction to the decay channel $H\rightarrow e^+e^-\mu^+\mu^-$.  }
	\label{fig:re_dia}
	\end{center}
\end{figure}

   In Fig.~\ref{fig:re_dia}, we have shown the real emission (photon) diagrams for the process $H\rightarrow e^+e^-\mu^+\mu^-$process. Each charged lepton can radiate a photon. As shown in Fig.~\ref{fig:re_dia}, there are $4$ real emission diagrams for the process $H\rightarrow e^+e^-\mu^+\mu^-$. 
    The photon is being emitted from the positron and the electron as shown in diagrams (a) and (b), respectively, in Fig.~\ref{fig:re_dia}. Similarly, photon radiation from anti-muon and muon are shown in diagrams (c) and (d) respectively in Fig.~\ref{fig:re_dia}.
    For the process $H\rightarrow 2e^+2e^-$, there are $8$ real emission diagrams. Diagrams are similar to those shown in Fig.~\ref{fig:re_dia} with crossing between electron and muon.
   The diagrams for this study have been generated using a {\tt Mathematica} package, {\tt FeynArts}~\cite{Hahn:2000kx}.

\section{Calculations}
\label{sec:calc_check}
  We compute a few hundred one-loop virtual diagrams along with tree-level, CT diagrams, and real emission diagrams.
  We treat leptons, light quarks as massless particles. A large set of diagrams is zero due to the vanishing coupling of scalars with the massless leptons and fermions.
   We ignore tadpole diagrams as they can be set to zero with the proper CT diagrams. With these considerations, we compute a total of $270$ and $540$ Feynman diagrams for the process $H\rightarrow e^+e^-\mu^+\mu^-$ and $H\rightarrow 2e^+2e^-$, respectively. 
   There are a few bubble-type diagrams for mixed type propagators between Goldstone boson and $\gamma/Z$ boson as the generic diagram (f) in Fig.~\ref{fig:loop_dia}. These diagrams do not contribute. 
    To calculate total virtual amplitudes, first we calculate the prototype type of diagrams, and then we map the rest of the diagrams to these diagrams. 
   The mapping is done using the proper choice of couplings, masses, and crossings of momenta.
  We calculate helicity amplitudes with spinor helicity formalism \cite{Peskin:2011in,Das:2023gsh}. 
   We use the t'Hooft-Veltman (HV) regularization scheme~\cite{THOOFT1972189} for these processes to calculate the virtual and CT diagrams.
   There are a few top-quark loop triangle diagrams in these processes. We follow the KKS scheme \cite{Shao:2011zza,Korner:1991sx} to calculate the trace of the fermion loop with $\gamma^5$.

   We use the symbolic manipulation program {\tt FORM}~\cite{Vermaseren:2000nd}, to simplify helicity amplitudes with helicity identities. Using {\tt FORM}, the amplitudes have been written in terms of spinor products and scalar products of vector objects with the momenta and polarizations.
    We calculate scalar integrals which appear at one loop using the package {\tt OneLOop}~\cite{vanHameren:2010cp}. The tensor integrals associated with the one-loop diagrams have been computed with an in-house code, {\tt OVReduce}~\cite{Agrawal:2012df,Agrawal:1998ch}.
    For tree and loop-level diagrams, we perform $4$-body phase-space integrals and for radiation diagrams, we perform $5$-body phase-space integrals. We use a Monte-Carlo integration package, called {\tt AMCI}~\cite{Veseli:1997hr} to perform the phase-space integrals. The package {\tt AMCI} is based on {\tt VEGAS} algorithm~\cite{Lepage:1977sw}. {\tt AMCI} uses Parallel Virtual Machine {\tt PVM}~\cite{10.7551/mitpress/5712.001.0001} for parallel computation across the  CPUs. We generate kinematic distributions with the help of the {\tt AMCI} routines.

\subsection{Renormalization and CMS scheme }
\label{subsec:renorm_cms}
In these processes, one-loop virtual diagrams are UV divergent. Rank-two, rank-three tensors are present in triangle-type diagrams, which lead to UV divergences. Bubble diagrams are also UV divergent.
To get the UV renormalized amplitude for these processes, we need to renormalize QED charge, vector boson masses, and the wave function of photon, $Z$, $W$-bosons, Higgs, and leptons. 
 The UV singularities have been removed with the appropriate CT diagrams shown in Fig.~\ref{fig:ct_dia}.
 As discussed in Sec .~\ref{sec:prcs}, the diagrams (a)-(e) in Fig.~\ref{fig:ct_dia} are the vertex CT diagrams, and the diagrams (f)-(i) are the self-energy CT diagrams.
   We have used the on-shell renormalization scheme~\cite{Denner:1991kt} for calculating the CT for these processes.
   To calculate counterterms, we need to calculate all self energy diagrms of photon, $Z$, $W$-bosons, Higgs, which also help us to scale a particular couplings in CT diagrams. 
 The CT diagrams given in Fig.~\ref{fig:ct_dia} remove all UV divergent poles from the desired one-loop virtual amplitudes. 
The unstable particles in the loop disturb perturbative computation when their off-shell loop momenta become on-shell near the threshold. To address this issue, we have implemented a complex mass scheme (CMS) \cite {DENNER200622} throughout the calculation.  In this scheme, the renormalization has been done in a modified version of the on-shell scheme to preserve the gauge invariance~\cite{Denner:2006ic,Denner:2005fg}. 
   As we calculate the self-energy diagrams, we can study the effect of anomalous $HHH$ coupling in counterterms. The Higgs boson self-energy diagrams are sensitive to anomalous $HHH$ coupling.
    The choice of input parameters for perturbative EW correction leads to the notion of the input parameter scheme. We have computed the partial decay width of Higgs in $G_F$ and $\alpha(M_Z)$ schemes. The running of $\alpha$ and charge renormalization has been taken appropriately in these two schemes \cite{Denner:1991kt,Andersen:2014efa,Eidelman:1995ny,PhysRevD.22.971}.

\subsection{IR divergences and dipole subtractions}
\label{subsec:ir_dp}
Most of the one-loop virtual amplitudes in this study are IR singular.
 There are real emission diagrams as shown in Fig.~\ref{fig:re_dia}, which are IR singular in soft and collinear regions. The virtual and real emission diagrams are separately singular, but their sum is finite.
  We have implemented the Catani-Seymour dipole subtraction scheme to remove IR singularities~\cite{Catani:1996vz}.  
 We followed the Ref.~\cite{Schonherr:2017qcj}, and implemented Catani-Seymour dipole subtraction for EW correction.
 There are $12$ dipole terms for each process. 
  The integrated dipole term in this scheme exactly cancels the IR poles from virtual amplitudes. The integrated dipole term also adds a finite contribution to the decay width.  The dipole terms behave the same as real emission amplitudes in collinear and soft regions. In the singular region, the large positive weight from real and large negative weights from dipoles together make the numerical integration evaluation convergent. For these processes, the dipole subtracted real emission contribution to the total partial decay width is at the permille level.

       \subsection{Anomalous couplings}
\label{subsec:anomalous_cpl}
\begin{figure}[!h]
  \begin{center}
\includegraphics [angle=0,width=1\linewidth]{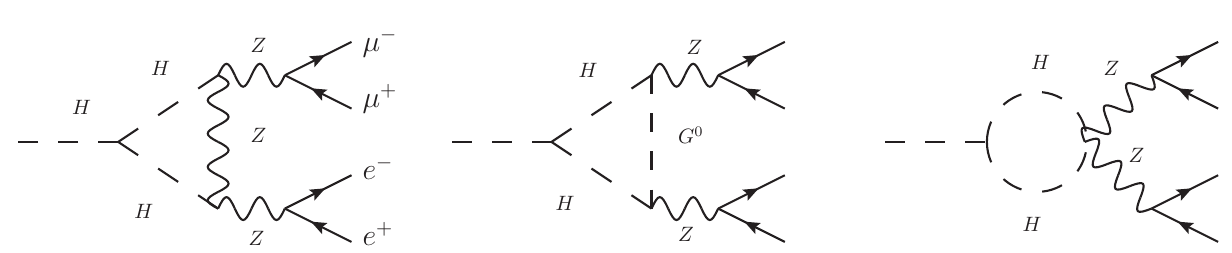}\\
	\caption{ NLO EW virtual diagrams with $HHH$ and $ZZHH$ couplings.  }
	\label{fig:hhh_dia}
	\end{center}
\end{figure}

\begin{figure}[!h]
  \begin{center}
\includegraphics [angle=0,width=0.8\linewidth]{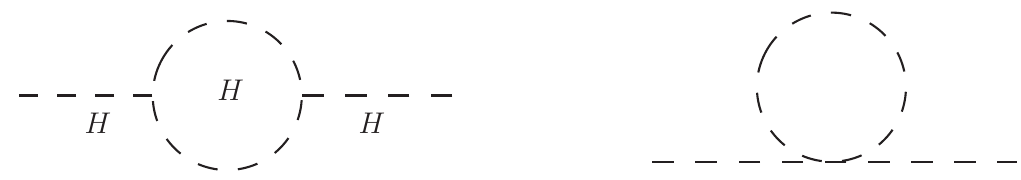}\\
	\caption{ Higgs self-enrgy  diagrams with $HHH$ and $ZZHH$ couplings.  }
	\label{fig:h_self}
	\end{center}
\end{figure}

  As we have discussed, these processes are sensitive to the anomalous behavior of $HHH$ couplings. As we have discussed in Sec.~\ref{sec:prcs} and~\ref{subsec:renorm_cms}, there are a few diagrams where we can introduce such anomalous coupling in $\kappa$-framework.
  The one-loop diagrams for the process $H\rightarrow e^+e^-\mu^+\mu^-$ where $HHH$ coupling is involved have been shown in Fig.~\ref{fig:hhh_dia}.
 There are three extra more diagrams for the process $H\rightarrow 2e^+2e^-$ which we have not shown. 
   $HHH$ coupling also appears in Higgs wave function renormalization that has been used in $HZZ$ vertex counterterm. Collectively the diagrams given in Fig.~\ref{fig:hhh_dia} are UV finite. 
   As the derivative of self enrgy digrams contributes to counterterms, the diagrams associated with the $HHH$ couplings in wave function renormalization of Higgs boson as shown in the Fig.~\ref{fig:h_self} (a), do not affect the UV renormalization. There is also $HHHH$ coupling in the Higgs self energy diagrams shown in Fig.~\ref{fig:h_self}(b). The derivative of this diagrams is zero, so we do not have any freedom to vary $HHHH$ coupling in these processes.  With this singular structure, the arbitrary scaling of $HHH$ do not affect the renormalizability in these processes.
  The $HHH$ coupling comes from the scalar potential term of the standard model Lagrangian and not from the gauge sector of the model. So, one can scale $HHH$ coupling independently with no loss of renormalizability and gauge invariance. 
  We scale $HHH$ coupling within the experimentally allowed region in the context of $\kappa$-framework and study its effect on the partial decay width of Higgs boson for these processes.


\section{Numerical Results}
\label{sec:numr_res}
We use the following SM parameter set for the computation of partial decay widths of Higgs.
\begin{gather}
 M_W=80.358\: {\text{GeV}},\:\:M_Z=91.153\: {\text{GeV}},\:\:M_t=172.5\: {\text{GeV}}\:\:M_H=125\: {\text{GeV}},\nonumber\\
\Gamma_W=2.0872 \: {\text{GeV}},\:\:\Gamma_Z=2.4944 \: {\text{GeV}},\:\Gamma_H=4.187\: {\text{MeV}} \:{\text{and}}\: \Gamma_t=1.481 \: {\text{GeV}}.
\label{eq:sm_param}
\end{gather}
We have treated light quarks and leptons as massless. The mass parameters and Weinberg angle have been made complex in context of the CMS scheme~\cite{DENNER200622}. We have calculated the decay widths in $G_F$ and $\alpha(M_Z)$ input parameter scheme. In $G_F$ scheme, the value of weak coupling can be calculated from the below relation 
\begin{equation}
\alpha_{G_F}=\frac{\sqrt{2}G_F{M_W}^2({M_Z}^2-{M_W}^2)}{\pi M_Z^2}.
\end{equation}
With the SM parameters given in Equ.~\ref{eq:sm_param}, the numerical value of weak coupling is  $\alpha=1/132.357$ in the $G_F$ scheme. The weak coupling value $\alpha=1/128.896$ in $\alpha(M_Z)$ scheme has been taken from the Ref.~\cite{Eidelman:1995ny}. We are also interested in kinematical differential distributions with respect to final state charged leptons. We put a minimum of $6$ GeV transverse momenta ($p_T$) cut on each charged lepton. 
We also isolate the same flavour lepton with the separation of $\Delta R=0.4$ in the $\eta-\phi$ plane. For the process $H\rightarrow 2e^+2e^-$, as there are same sign leptons in the final state, we make $ p_T$-ordered leptons. The leading (highest $p_T$) and sub-leading lepton has been denoted by subscript $1$ and $2$ (i.e., $e_1$, $e_2$) respectively.
As the contribution from the dipole subtracted real emission amplitude is at the permille level, we ignore this contribution for kinematical distributions.
For the total decay width computation, we have not implemented any cuts on final state leptons. 
\subsection{SM prediction}
\label{subsec:sm_pdct}

   We have listed the standard model prediction for the partial decay width of Higgs boson for the process $H\rightarrow e^+e^-\mu^+\mu^-$ in Tab.~\ref{table:sm_rst1}  and for the process $H\rightarrow 2e^+2e^-$ in Tab.~\ref{table:sm_rst2}  in the $G_F$ and $\alpha(M_Z)$ schemes. We define relative enhancement as $\text{RE}=\frac{\Gamma^{NLO}-\Gamma^{LO}}{\Gamma^{LO}}\times 100\:\%$, where $\Gamma^{LO}$ and $\Gamma^{NLO}$ are leading order (LO) and next-to-leading order (NLO) decay widths respectively. The LO partial decay width are $238.04$ eV and $256.82$ eV; whereas the NLO (EW) corrected decay widths are $241.03$ eV and $237.69$ eV in the $G_F$ and $\alpha(M_Z)$ schemes, respectively, for the process $H\rightarrow e^+e^-\mu^+\mu^-$.
    For this process, the relative enhancement in the $G_F$ scheme is $1.26\%$, whereas in the $\alpha(M_Z)$ scheme the relative enhancement is $-7.45\%$.
    The LO partial decay width are $131.51$ eV and $142.39$ eV; whereas the NLO (EW) corrected decay widths are $131.84$ eV and $129.76$ eV in the $G_F$ and $\alpha(M_Z)$ schemes respectively for the process $H\rightarrow 2e^+2e^-$.
    For this process the relative enhancement in the $G_F$ scheme is $0.25\%$, whereas in the $\alpha(M_Z)$ scheme, the relative enhancement is $-8.87\%$.
    
     Although the LO results differ quite significantly but the NLO (EW) corrected results differ by only $\sim 1.5\%$ for the two input parameter schemes. Our results are having good agreement with the {\tt Prophecy4f} package~\cite{BREDENSTEIN2006131,Bredenstein:2006nk}.
       As we can see from the Tab.~\ref{table:sm_rst1} and ~\ref{table:sm_rst2}, the relative enhancement is smaller in the $G_F$ scheme, i.e., in the lower-order prediction, the universal correction has been incorporated.
\begin{table}[H]
\begin{center}
\begin{tabular}{|c|c|c|c|}
\hline
Input&&&\\
parameter&$\Gamma^{LO}$ (eV)&$\Gamma^{NLO}$ (eV)&RE\\
scheme&&&\\
\hline
&&&\\
$G_F$&$238.04$&$241.03$&$1.26\%$\\
&&&\\

\hline
&&&\\
$\alpha(M_Z)$&$256.82$&$237.69$&$-7.45\%$\\
&&&\\
\hline
\end{tabular}
\caption{Partial decay widths of Higgs boson in the channel $H\rightarrow e^+e^-\mu^+\mu^-$ in the $G_F$ and $\alpha(M_Z)$ schemes and their relative enhancements.}
\label{table:sm_rst1}
\end{center}
\end{table}
\begin{table}[H]
\begin{center}
\begin{tabular}{|c|c|c|c|}
\hline
Input&&&\\
parameter&$\Gamma^{LO}$ (eV)&$\Gamma^{NLO}$ (eV)&RE\\
scheme&&&\\
\hline
&&&\\
$G_F$&$131.51$&$131.84$&$0.25\%$\\
&&&\\

\hline
&&&\\
$\alpha(M_Z)$&$142.39$&$129.76$&$-8.87\%$\\
&&&\\
\hline
\end{tabular}
\caption{Partial decay widths of Higgs boson in the channel $H\rightarrow 2e^+2e^-$ in the $G_F$ and $\alpha(M_Z)$ schemes and their relative enhancements.}
\label{table:sm_rst2}
\end{center}
\end{table}

\begin{figure}[htp]
\begin{center}
\includegraphics [angle=0,width=0.47\linewidth]{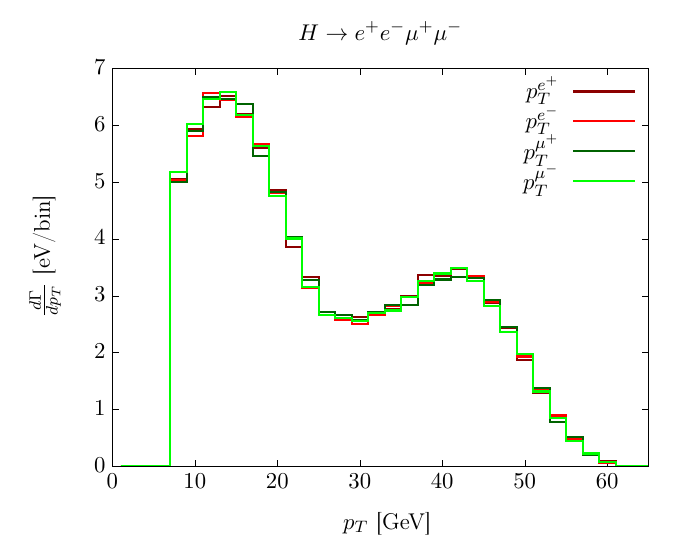}
\includegraphics [angle=0,width=0.47\linewidth]{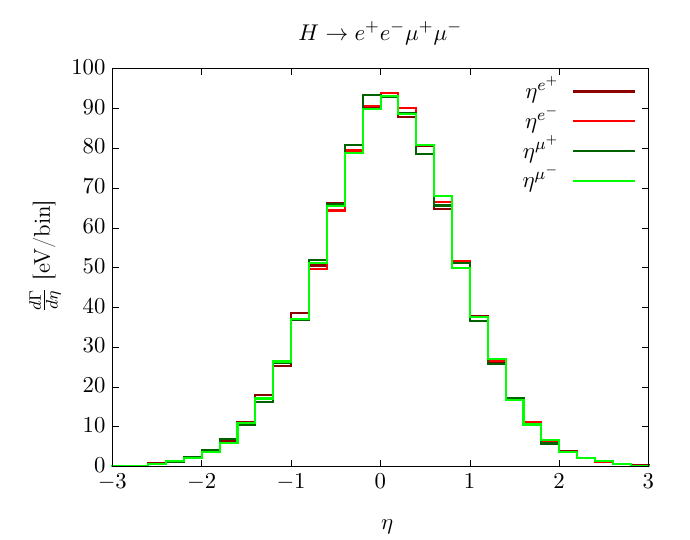}
\includegraphics [angle=0,width=0.57\linewidth]{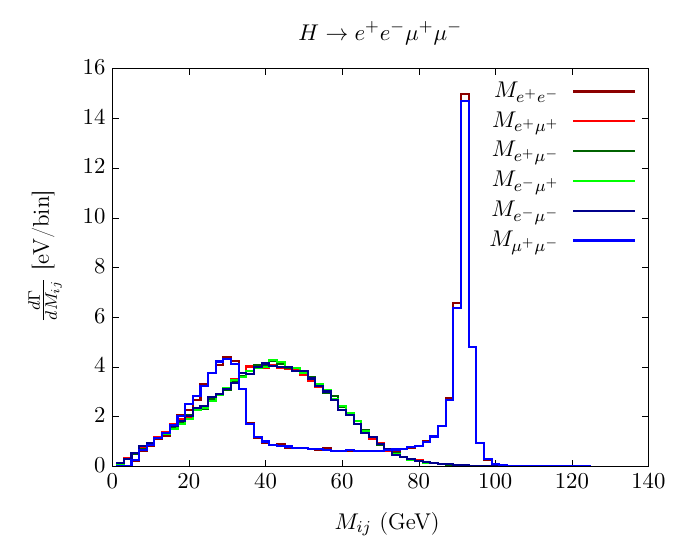}
\caption{ Transverse momenta ($p_T$), rapidity ($\eta$)  and invariant-mass distribution ($M_{ij}$) for the process $H\rightarrow e^+e^-\mu^+\mu^-$.}
\label{fig:dis_smme}
\end{center}
\end{figure}

\begin{figure}[htp]
\begin{center}
\includegraphics [angle=0,width=0.47\linewidth]{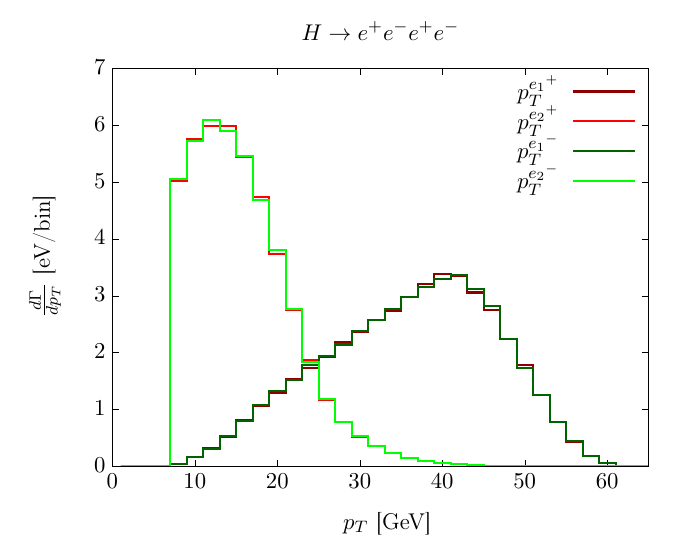}
\includegraphics [angle=0,width=0.47\linewidth]{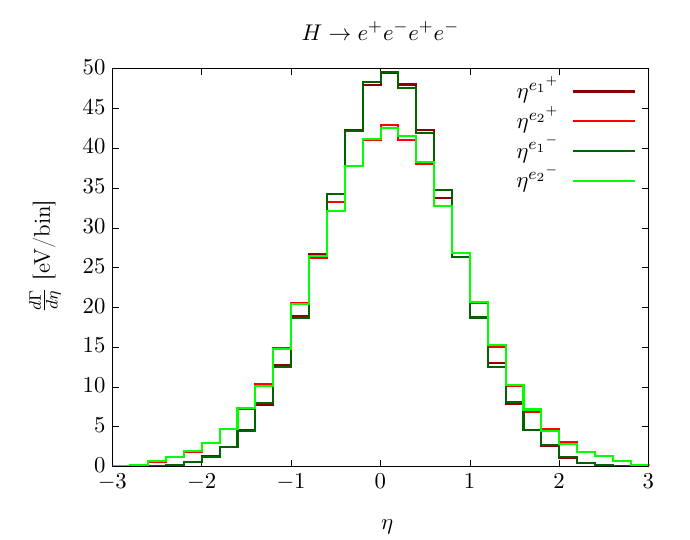}
\includegraphics [angle=0,width=0.57\linewidth]{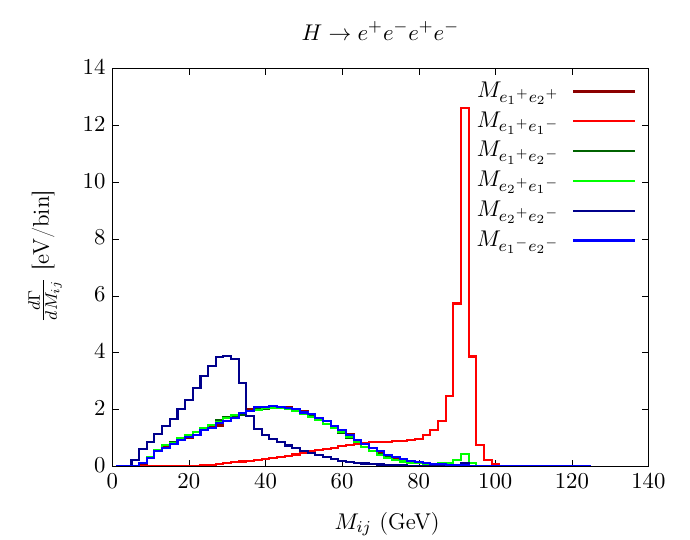}
\caption{ Transverse momenta ($p_T$), rapidity ($\eta$)  and invariant-mass distribution ($M_ij$) for the process $H\rightarrow 2e^+2e^-$.}
\label{fig:dis_smee}
\end{center}
\end{figure}

In Fig \ref{fig:dis_smme}, we have plotted the $p_T$,
$\eta$, and mass distributions for the leptons including
electroweak corrections for the process $H\rightarrow e^+e^-\mu^+\mu^-$. In Fig \ref{fig:dis_smee}, the same distributions are plotted for the process $H\rightarrow 2e^+2e^-$ w.r.t. leading ($e_1$) and sub-leading ($e_2$) leptons. These distributions are similar to those without
electroweak corrections. The leptons are central with expected $p_T$ and mass distributions. The $p_T$ of harder leptons are peaked near $M_Z/2$, and softer to a lower value. The mass distributions are also peaked at $M_Z$ and
around $35$ GeV.

\subsection{Anomalous coupling effect}
\label{subsec:anoma_cpl_efct}
  We vary $\kappa$ of $HHH$ coupling within the experimental bounds and study their effects on the decay width $\Gamma^{NLO}$. 
   We define relative increment as ${\text{RI}}=\frac{\Gamma^{NLO}_\kappa-\Gamma^{NLO}_{SM}}{\Gamma^{NLO}_{SM}}\; \times 100\%$.
   
   We vary $\kappa_{HHH}$ from $10$ to $-10$~\cite{TheATLAScollaboration:2014scd} and enlisted corresponding relative increment in Tab.~\ref{table:hhh_kappa_p1} and Tab.~\ref{table:hhh_kappa_p2} for the process $H\rightarrow e^+e^-\mu^+\mu^-$ and  $H\rightarrow 2e^+2e^-$ respectively. As we see from the Tab.~\ref{table:hhh_kappa_p1} for the process $H\rightarrow e^+e^-\mu^+\mu^-$, the RI goes from $0.36\%$ to $-23.91\%$ in the $G_F$ scheme depending on $\kappa_{HHH}$ value and becomes positive near $\kappa_{HHH}\sim 2\text{-}4$; 
   whereas in the $\alpha(M_Z)$ scheme, the RI goes from $0.40\%$ to $-26.72\%$ and become positive near $\kappa_{HHH}\sim 2\text{-}4$.
   A similar behavior is seen for the process $H\rightarrow 2e^+2e^-$ in the Tab.~\ref{table:hhh_kappa_p2}, the RI goes from $0.01\%$ to $-24.60\%$ in the $G_F$ scheme depending on $\kappa_{HHH}$ value and becomes positive near $\kappa_{HHH}\sim 2\text{-}4$; 
   whereas in the $\alpha(M_Z)$ scheme, the RI goes from $0.26\%$ to $-27.26\%$ and become positive near $\kappa_{HHH}\sim 2$.
\begin{table}[H]
\begin{center}
\begin{tabular}{|c|c|c|}
\hline
\multirow{2}{*}{$\kappa_{HHH}$}&\multicolumn{2}{|c|}{RI}\\
\cline{2-3}
&$G_F$ scheme&$\alpha(M_Z)$ scheme\\
\hline
$10$&$-7.65$&$-8.62$\\
\hline
$8$&$-3.84$&$-4.32$\\
\hline
$6$&$-1.23$&$-1.39$\\
\hline
$4$&$0.17$&$0.19$\\
\hline
$2$&$0.36$&$0.40$\\
\hline
$-1$&$-1.62$&$-1.83$\\
\hline
$-2$&$-3.17$&$-3.25$\\
\hline
$-4$&$-6.31$&$-7.12$\\
\hline
$-6$&$-10.95$&$-12.41$\\
\hline
$-8$&$-16.82$&$-18.95$\\
\hline
$-10$&$-23.91$&$-26.72$\\
\hline
\end{tabular}
\caption{Effect of anomalous $HHH$ coupling on the partial decay width of the process $H\rightarrow e^+e^-\mu^+\mu^-$.}
\label{table:hhh_kappa_p1}
\end{center}
\end{table}
\begin{table}[H]
\begin{center}
\begin{tabular}{|c|c|c|}
\hline
\multirow{2}{*}{$\kappa_{HHH}$}&\multicolumn{2}{|c|}{RI}\\
\cline{2-3}
&$G_F$ scheme&$\alpha(M_Z)$ scheme\\
\hline
$10$&$-8.46$&$-8.90$\\
\hline
$8$&$-4.32$&$-4.60$\\
\hline
$6$&$-1.68$&$-1.66$\\
\hline
$4$&$-0.20$&$0.05$\\
\hline
$2$&$0.01$&$0.26$\\
\hline
$-1$&$-2.08$&$-2.03$\\
\hline
$-2$&$-3.38$&$-3.41$\\
\hline
$-4$&$-6.77$&$-7.38$\\
\hline
$-6$&$-11.29$&$-12.91$\\
\hline
$-8$&$-17.19$&$-19.48$\\
\hline
$-10$&$-24.60$&$-27.26$\\
\hline
\end{tabular}
\caption{Effect of anomalous $HHH$ coupling on the partial decay width of the process $H\rightarrow 2e^+2e^-$.}
\label{table:hhh_kappa_p2}
\end{center}
\end{table}

\begin{figure}[htp]
\begin{center}
\includegraphics [angle=0,width=0.47\linewidth]{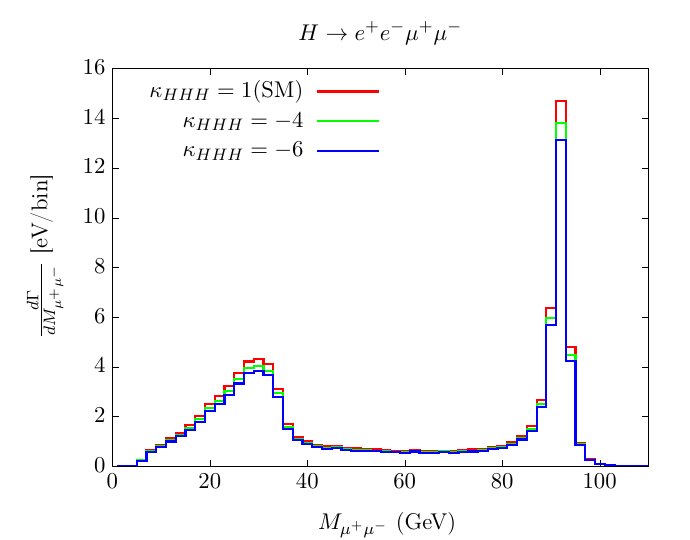}
\includegraphics [angle=0,width=0.47\linewidth]{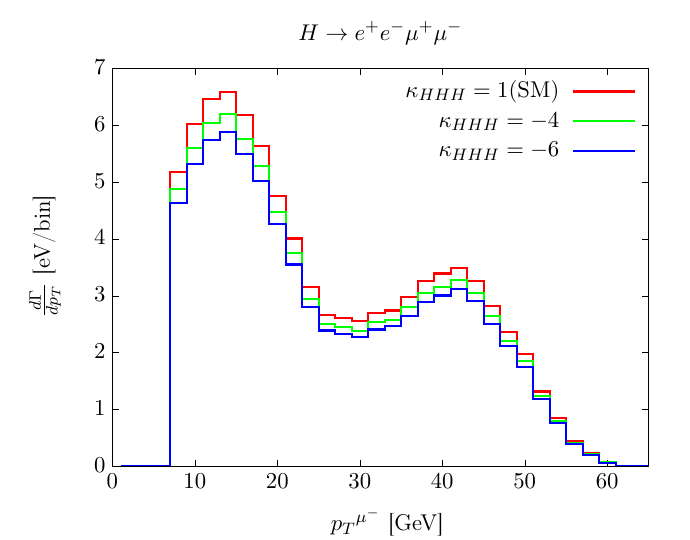}
\caption{ Transverse momenta distributions for the process $H\rightarrow e^+e^-\mu^+\mu^-$ with $\kappa_{HHH}=-4,-6$.}
\label{fig:dis_mepT}
\end{center}
\end{figure}

\begin{figure}[htp]
\begin{center}
\includegraphics [angle=0,width=0.47\linewidth]{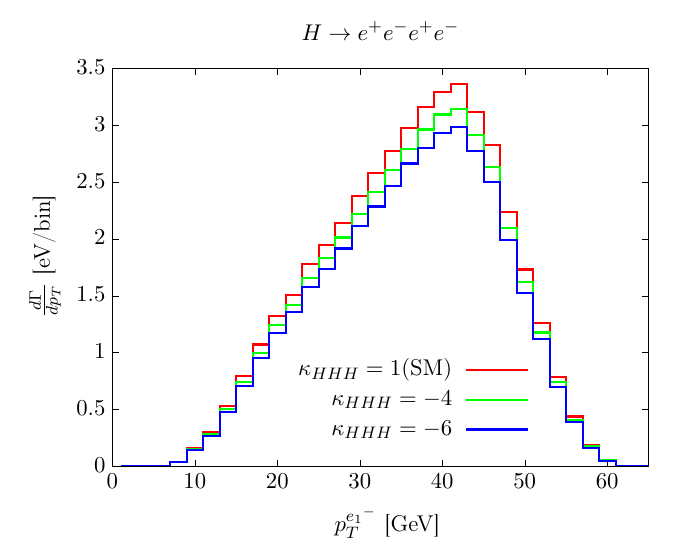}
\includegraphics [angle=0,width=0.47\linewidth]{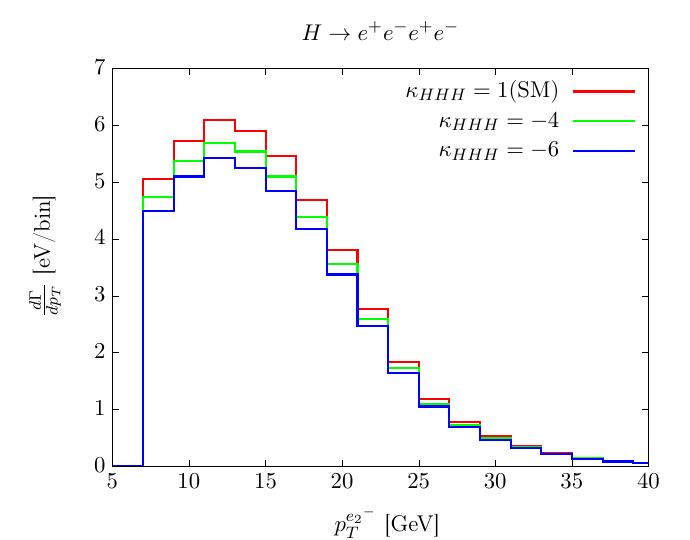}
\caption{ Transverse momenta distribution (leading ) for the process $H\rightarrow 2e^+2e^-$ with $\kappa_{HHH}=-4,-6$.}
\label{fig:dis_eepT}
\end{center}
\end{figure}

\begin{figure}[htp]
\begin{center}
\includegraphics [angle=0,width=0.47\linewidth]{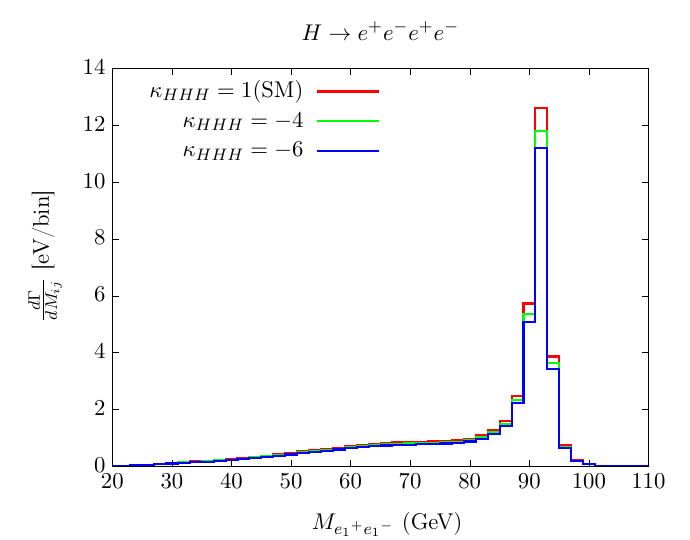}
\includegraphics [angle=0,width=0.47\linewidth]{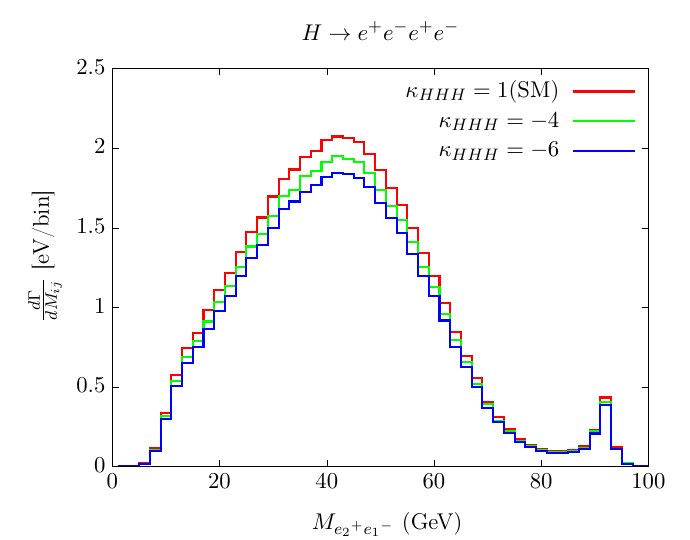}
\caption{ Invariant-mass distributions ($M_{ij}$) for the process $H\rightarrow 2e^+2e^-$ with $\kappa_{HHH}=-4,-6$.}
\label{fig:dis_eem}
\end{center}
\end{figure}

In Fig.~\ref{fig:dis_mepT}, we 
have displayed the effect of anomalous coupling on the $p_T$
and mass distributions of the muons in the process $H\rightarrow e^+e^-\mu^+\mu^-$. In Figs.~\ref{fig:dis_eepT} and ~\ref{fig:dis_eem} same distributions are displayed for the process $H\rightarrow 2e^+\,2e^-$.
As for the partial width, we see that more negative is $\kappa_{HHH}$, more negative is
the corrections. The effect of the scaling is more visible
near the peak of the distributions. In putting a bound on the $HHH$ coupling, a fit to the distributions may help us
in putting a tighter bound on the coupling.
\section{Conclusion}
\label{sec:conclusion}
In this letter,
 we have studied the one-loop EW correction to the process $H\rightarrow Z\,Z^*\rightarrow$ 4 charged leptons. Our results for the standard model electroweak corrections are in agreement with the previous calculation. We investigate the effect of the anomalous $HHH$ coupling on the partial decay width and decay distributions of the Higgs boson. We have recomputed a few vertex diagrams, and Higgs boson wave function renormalization constant to include the effects of
 the scaling of the $HHH$ coupling. We varied the $\kappa_{HHH}$ and observe a significant change in the decay width. The behavior of RI on varying $\kappa_{HHH}$ is the same for the two input schemes, as the scaling of $HHH$ coupling does not affect gauge invariance. We have also presented the effect of varying
 the $HHH$ coupling on the decay distributions. For $\kappa =-4$, the partial decay width and the distributions can change by about $-6\%$.
 This change can be observable at 
 the HL-LHC, and be used to put a tighter bound on the $HHH$ coupling and rule out a few alternate Higgs sector scenarios.

\bibliographystyle{JHEP}
\bibliography{h24clref}

\end{document}